\documentstyle[aps,epsf]{revtex}

\begin{document}


\title{BCS vs Overhauser pairing in dense (2+1)d QCD}

\author{Prashanth Jaikumar and Ismail Zahed} \address{ Department of
Physics \& Astronomy, SUNY at Stony Brook, New York 11794-3800}
\date{\today} 
\maketitle

\begin{abstract}

We compare the BCS and Overhauser effect as competing mechanisms for the
destabilization of the quark Fermi surface at asymptotically large
chemical potential, for the special case of 2 space and 1 time
dimension. We use the framework of perturbative one-gluon
exchange, which dominates 
the pairing at $\mu/g^2\gg 1$. With screening in matter, we show that in
the weak 
coupling limit the Overhauser effect can compete with the BCS effect
only for a sufficiently 
large number of colors. Both the BCS and the Overhauser gaps are of
order $g^{4}/\mu$ in Landau gauge.

\vskip 0.2cm
PACS number(s): 12.38Aw 11.30Er 26.60.+c
\end{abstract}


 \section{Introduction}

Quantum Chromodynamics (QCD) at high density was first studied in the 
late seventies~\cite{Seventies,LOVE} and has received renewed
attention, especially in the last three years~\cite{Renew,ALL98,ALL99,ALL00}. 
At large baryon density and small temperature, it
is relevant to studies of neutron stars and heavy ion collisions
in the baryon rich regime. Theoretical studies of dense quark matter
have revealed the preferred phase to be a color superconductor with
several novel features 
like color-flavor locking, chiral symmetry breaking, 
color Meissner effect and the formation of a mass gap.

\vskip 0.3cm 

     At asymptotic densities, the Fermi surface for quarks is
well defined and the low energy excitations are weakly interacting
quasiparticles and holes. In QCD, perturbative 1- gluon exchange can
provide an arbitrarily weak attractive interaction in the color anti-symmetric
${{\bf{\bar{3}}}}$ channel that destabilizes the Fermi surface and causes
particle-particle pairing. Another mechanism of destabilization is
particle-hole pairing\,(Overhauser effect)~\cite{ove}, studied earlier
in a variational method and for large ${N_{c}}$~\cite{dgr,schu}. More
recently, it has been shown that the equations driving the
particle-hole instability resemble those of the BCS instability, modulo
phase-space factors~\cite{rho}. A novel mechanism for destabilizing the 
Fermi surface through a BCS crystal was also recently suggested
in~\cite{KRISHNA} and may also compete with the instabilities to
be studied below. Its consideration is outside the scope of this work.

\vskip 0.3cm

Screening affects the strengths of the BCS and Overhauser pairing. The authors of~\cite{rho} have shown that in the regime of weak coupling
in 3+1d, the BCS effect is dominant upto a large number of colors.
In 1+1d, the drastic reduction in phase space for BCS pairing
with no comparable reduction for Overhauser pairing means that the
latter dominates whatever ${N_{c}}$. In this work, we wish to establish the case for 2+1d, taking into account the screening in this space-time dimension. 

\vskip 0.3cm

This paper is organized as follows. 
In section {\bf II}, we derive the quark 1-loop contribution to the gluon propagator in a covariant gauge, at finite temperature and
density (T${\ll \mu}$ for cold and dense matter). 
Section {\bf III} motivates the symmetry of the particle-particle and particle-hole condensates in 2+1 dimensions, followed by formally differing derivations of the gap equations for these two objects. The resulting gap equations are solved in the presence of electric screening and Landau damping in dense quark matter. We present our results for the gaps in the Landau gauge. The conclusions of our analysis are in section {\bf IV}.



\section{Screening in 2+1d}

 We begin by evaluating the gluon polarization tensor in matter in the
imaginary time formalism~\cite{han}.
\begin{equation}
\Pi^{\mu\nu}(q_{0},\bf{\vec{q}})\rm = {\it{g}}^2
{\it{T}\sum_{n}}\int\frac{{\it{d}}^{\rm{2}}
{\it{p}}}{(2\pi)^{\rm{2}}}{\rm{Tr}}\left(\gamma^{\mu}\frac{1}{{\it{p}}{\hskip-2.0mm}/}
\gamma^{\nu}\frac{1}{{\it{p}}{\hskip-2.0mm}/ + {\it{q}}{\hskip-2.0mm}/}\right) \quad,
 \end{equation}

where $q$ , $p$ and $p+q$ denote respectively the 4-momentum of the gluon and  the two internal quark lines. At finite temperature and density, one makes the usual identification ${ip_{3} = p_{0} = (2n+1)i\pi T + \mu}$ for fermions and ${iq_{3} = q_{0} = 2n i\pi T}$ for bosons (here, $n$ labels the Matsubara frequency). In 2+1d, the Pauli matrices $\gamma_\mu\equiv\sigma_{\mu}\equiv (\sigma_3,i\vec{\sigma})$ generate a representation of the Clifford algebra ${\{\gamma_{\mu}\gamma_{\nu}\} = 2g_{\mu\nu}}$ with
${g_{\mu\nu} = \rm diag(1,-1,-1)}$. In an arbitrary covariant gauge with gauge parameter $\lambda$, the inverse propagator in matter ${(D^{-1})^{\mu\nu}}$ and ${\Pi^{\mu\nu}}$ are related as \cite{kap} 

\begin{equation}
(D^{-1})^{\mu\nu} = q^2 g^{\mu\nu} + \Pi^{\mu\nu} - q^{\mu\nu}\left(1-\frac{1}{\lambda}\right)  \quad. 
\end{equation}

 Evaluating the Matsubara sum using
contour integration, omitting the vacuum piece (T and ${\mu}$ -
independent), and analytically continuing ${iq_{3} = \omega +
i\epsilon}$, we get in the limit ${T \rightarrow 0}$ (with
\bf{q} ${= \mid\vec{\bf{q}}\mid}$)

\begin{eqnarray}
\Pi^{00} &=& {\it g}^{\rm 2}\rm
\left[\hspace{0.1in} \left(-\frac{\mu}{2\pi} +\frac{\bf{q}\rm}{16} \right) + i
\frac{\omega}{\bf{q}\rm}\left(\frac{-\mu}{4\pi}\right)\hspace{0.1in}
\right]\hspace{0.1in} \theta (\bf{q}\rm-\omega) +{\it g}^{\rm 2}\rm
\left[\hspace{0.1in}\frac{\bf{q}\rm^2}{\omega^2} \left(\frac{\mu}{4\pi}\right)  + i
\frac{\bf{q}\rm^2}{\omega^2}\left(\frac{-\omega}{16}\right)\hspace{0.1in}
\right]\hspace{0.1in} \theta (\omega-\bf{q}\rm)  \label{Pi00}\\
\Pi_{\perp} &=& {\it g}^{\rm 2}\rm
\left[\hspace{0.1in} \left(\frac{\bf{q}\rm}{16} -
\frac{\omega^2}{\bf{q}\rm^2}\left(\frac{\mu}{2\pi}\right)\right) + i
\frac{\omega}{\bf{q}\rm}\left(\frac{3\mu}{8\pi}\right)\hspace{0.1in}
\right]\hspace{0.1in} \theta (\bf{q}\rm-\omega) +{\it g}^{\rm 2}\rm
\left[\hspace{0.1in} \left(-\frac{\mu}{4\pi}\right) - i\left(
\frac{\mu}{2\pi}+\frac{\omega}{16}\right)\hspace{0.1in} \right]\hspace{0.1in}
\theta (\omega-\bf{q}\rm)  \label{Piperp}\\
\Pi_{\bf{L}} &=&  {\it g}^{\rm 2}\rm
\left[\hspace{0.1in} \frac{\omega^2}{\bf{q}\rm^2}\left(\frac{\mu}{2\pi}\right) + i
\frac{\omega^3}{\bf{q}\rm^3}\left(\frac{\mu}{4\pi}\right)\hspace{0.1in}
\right]\hspace{0.1in} \theta (\bf{q}\rm-\omega) + {\it g}^{\rm 2}\rm
\left[\hspace{0.1in} \left(-\frac{\mu}{4\pi}\right) + i\left(
\frac{\omega}{16}\right)\hspace{0.1in} \right]\hspace{0.1in} \theta
(\omega-{\bf{q}}) \label{Pilong}\quad.
\end{eqnarray}
\vskip 0.2cm
 \rm where ${\Pi^{ij} = \Pi_{\perp} {\rm{P}_{\perp}}^{ij} + \Pi_{\bf{L}}
{\rm{P}}_{\bf{L}}^{ij}}$, with ${{\rm{P}_{\perp}}^{ij} = (\delta^{ij} -
\frac{q^{i}q^{j}}{{\bf{q}}^2})}$  and ${{\rm{P}}_{\bf{L}}^{ij} =
\frac{q^{i}q^{j}}{q^2}}$ ($i,j$=1 to 3).  

In deriving Eqs.~(\ref{Pi00}-\ref{Pilong}), the approximations\footnote{The gaps in 2+1d are power law suppressed, and care is needed in introducing this approximation in our perturbative result.} ${\omega,\bf{q}\rm \ll \mu}$ have been made since energy transfers in the scattering of $qq$ (BCS) and $\bar{q}q$ (Overhauser) pairs are of the order of the gap, and momentum transfers can be taken to be small for forward scattering of quarks. In Euclidean space, the screened gluon propagator in
Feynman gauge reads

\begin{equation}
 D_{E,M}(x-y) = \int \frac{d^3
 q}{(2\pi)^3}\frac{e^{-iq.(x-y)}}{q^2+m^{2}_{E,M}} \quad  ,
\end{equation}
where ${m^{2}_{E} = m^{2}_{D}}$ refers to the Debye mass and ${m^{2}_{M}}$ to
the magnetic scale generated by Landau damping. Their values
(generalized to ${N_{f}}$ flavors) follow from (3) and (4) respectively and are
given by
$m^{2}_{E}/g^2\mu = m^{2}_{D}/g^2\mu =
N_{f}/2\pi$ and $m^{2}_{M}/m^{2}_{D} =
\frac{3}{4}|q_{3}|/{\bf{q}}$. Note that the coupling
${g^{2}}$ has mass dimensions. The results are similar to that in 3+1d
for which $m^{2}_{E}/g^2\mu^2 = m^{2}_{D}/g^2\mu^2 =
N_{f}/2\pi^2$ and $m^{2}_{M}/m^{2}_{D} =
\frac{\pi}{4}|q_{4}|/{\bf{q}}$.



\section{Condensates and Gap Equations in 2+1d}

\subsection{\underline{Condensates}}
 
  Usually, diquarks are taken to condense in the parity even
${0^{+}}$ channel. In 2+1d, parity is defined by
inversion about 1 spatial coordinate  only, as it is an improper
transformation. For the sake of definiteness, we define the
parity transformation in space as ${x_{1}\rightarrow
x_{1},\hspace{0.05in} x_{2}\rightarrow{-x_{2}}}$. The parity operator
in spin space is then ${\sigma_{1}}$ which implies that a term like
${\bar{q} q}$ is not invariant under this transformation. Thus, for
the particle-hole pseudoscalar condensate with 2 flavors, we have the action of the parity operator (P) as 

\begin{equation}
\bar{q} {\bf{1}}^{c} {\bf{1}}^{f} q = \bar{u} u + \bar{d} d
\stackrel{\mbox{P}}{\rightarrow} -\bar{u} u - \bar{d} d \quad ,
\end{equation}
where $c$ and $f$ denote color and flavor respectively.
A scalar condensate may still be written down if we allow parity to
change isospin as ${ u(d)\stackrel{\mbox{P}}{\rightarrow} +/-
\hspace{0.05in}\sigma_{1} d(u)}$. Then we have

\begin{equation}
\bar{q}{\bf 1}^{c} \sigma_{3}^{f} q = \bar{u} u - \bar{d} d
\stackrel{\mbox{P}}{\rightarrow} -\bar{d} d + \bar{u} u  \quad .
\end{equation}
The ${+/-}$ indicates that there are 4 possible sign combinations we
can choose in the combined transformation of the u and d quarks. Note
that this new definition of parity does not affect the way the
pseudoscalar transforms. Now, we will see that the relative sign is
also determined for us by the particle-particle condensates. 

\vskip 0.3cm

The charge conjugated spinor transforms under parity as ${
q^{c}\stackrel{\mbox{P}}{\rightarrow} \mbox{det}(\Lambda_{\mbox{P}}) \sigma_{1} q^{c}}$, where
${\Lambda_{\mbox{P}}}$ denotes the matrix corresponding to the improper Lorentz transformation P. There is an extra minus sign as compared to the parity transformation of the quark spinor. The charge conjugation matrix ${C}$ is ${\sigma_{2}}$, which is antiymmetric. The attractive color channel $\bar{{\bf{3}}}$ is
anti-symmetric and the overall wave function for the BCS pairing must
also be anti-symmetric, therefore, one is forced to choose ${\sigma_{ 2}^{f}}$
for the flavor matrix. Then, the scalar BCS
condensate is

\begin{equation}
\bar{q}^{c} \lambda_{2}^{c}\,\sigma_{2}^{f} q = \bar{u}^{c}d -
{\bar{d}}^{\,c}u \quad .
\end{equation}
For this to be even-parity, both $u$ and $d$ should have the same sign in
their transformation. On the other hand, a pseudoscalar BCS condensate will
necessarily have a relative sign between the transformations for  $u$ and $d$. In this paper, it is the scalar which is studied.

\vskip 0.3cm

These problems arise in odd space-time dimensions due to the lack of a consistent definition for a gamma matrix that anti-commutes with all the ${\gamma_{\mu}}$'s, namely ${\gamma_{5}}$. We proceed now to the gap equations for the scalar particle-paticle and particle-hole condensates. 

\vskip 0.5cm 


\subsection{\underline{ BCS Gap equations}}

 In order to derive the BCS gap equation, we will use
the standard Nambu-Gorkov formalism, which introduces
charge-conjugated spinors to enable the writing of the generating
functional of QCD at finite chemical potential in a form identical to
that at zero chemical potential~\cite{ris}. With charge conjugation
defined as ${\psi^{c} = C{\bar{\psi}}^{T}}$, the Nambu-Gorkov spinor is ${\Psi = (\psi,\bar{\psi}^T)}$ and can be expressed as a
4$. N_{c}.N_{f}$ column vector. Neglecting quark mass effects, the
form of the gap matrix is
\begin{equation}
\Delta_{ij}^{ab}(q) = (\lambda_{2})^{ab}(\sigma_{2})_{ij}
C\left\{\Delta_{1}(q)\frac{1}{2}(1+\vec{\sigma}\times\hat{q}) +
\Delta_{2}(q)\frac{1}{2}(1-\vec{\sigma}\times\hat{q})\right\} \label{ansatz}   \quad.
\end{equation}  
where
$\Lambda_{\pm} = \frac{1}{2}(1\pm \vec{\sigma}\times\hat{q})
$
are the particle and anti-particle projectors respectively in 2+1d. (In
the weak coupling limit $\alpha/\mu\ll 1$ with $\alpha=g^2/4\pi$,
 the equations of motion allow us to replace
${\vec{\sigma}\times\hat{q}}$ by the unit matrix so that only the gap
${\Delta_{1}}$ remains and the anti-gap ${\Delta_{2}}$ drops
out. Moreover, the anti-gap is strongly gauge dependent.) The self-energy in the
Nambu-Gorkov formalism obeys the Schwinger-Dyson (SD) equation
\begin{equation}
\Sigma(k) =
-ig^{2}\int\frac{d^{3}q}{(2\pi)^3}\Gamma_{\mu}^{a}\,S(q)\,D_{\mu\nu}^{ab}(q-k)\,\Gamma_{\nu}^{b}   \quad.
\end{equation}
Here, ${\Sigma(k) = S^{-1}(k) - S_{0}^{-1}(k)}$ is the proper
self-energy with $S(k)$ as defined in ~\cite{IZ}. ${\Gamma_{\mu}^{a}}$ is the quark-gluon vertex which
we take to be the leading order result in perturbation theory,
\begin{equation}
\Gamma_{\mu}^{a} = \biggl(\begin{array}{cr} \gamma_{\mu}\lambda^{a}/2
& 0 \\ 0 & -(\gamma_{\mu}\lambda^{a}/2)^{T}
\end{array}\biggr)  \quad.
\end{equation}
We also neglect the wave-function renormalization to leading order
in the weak coupling so that the diagonal part of the self-energy has
only the free quark propagators. Therefore,
\begin{equation}
S^{-1}(q) = \biggl(\begin{array}{cr}
q{\hskip-2.0mm}/+\mu{\hskip-2.0mm}/ & \bar{\Delta} \\ \Delta &
(q{\hskip-2.0mm}/-\mu{\hskip-2.0mm}/)^{T}
\end{array}\biggr) \label{invprop}   \quad,
\end{equation}
where ${\bar{\Delta} =
\gamma_{0}\Delta^{\dag}\gamma_{0}}$. Determining ${S_{21}}$ from Eq.~(\ref{invprop}), inserting the ansatz Eq.~(\ref{ansatz}) for the gap, and projecting out the positive energy part, we obtain the integral equation
\begin{eqnarray}
\Delta_{1}(k) =
-\frac{N_{c}+1}{2N_{c}}ig^{2}\int\frac{d^{3}q}{(2\pi)^3}\biggl[\frac{\Delta_{1}(q)}{q_{0}^{2}-(\mid\vec{\bf{q}}\mid
- \mu)^{2} - \Delta_{1}(q)^{2}}\biggl\{\biggl(\frac{1}{2}+\frac{1}{2}\hat{k}\cdot\hat{q}\biggr)D_{\perp}(k-q)\nonumber \\ 
+ \biggl(\frac{1}{2}+\frac{1}{2}\hat{k}\cdot\hat{q}\biggr)D_{L}(k-q)+\biggl(K^{0^2}(1+\hat{k}\cdot\hat{q})\biggr)\frac{D_{\lambda}}{2}\biggr\} \nonumber \\ 
 + \frac{\Delta_{2}(q)}{q_{0}^{2}-(\mid\vec{\bf{q}}\mid + \mu)^{2} -
\Delta_{2}(q)^{2}}\biggl\{\biggl(1-\hat{k}\cdot\hat{q}\biggr)D_{\perp}(k-q) \nonumber \\
 + \biggl(\frac{1}{2}-\frac{1}{2}\hat{k}\cdot\hat{q}\biggr)D_{L}(k-q) + 
\biggl(K^{0^2}(1-\hat{k}\cdot\hat{q})-|\vec{\bf{K}}|^{2}(1+\hat{k}\cdot\hat{q})\biggr)\frac{D_{\lambda}}{2}\biggr\}\biggr]   \quad,
\end{eqnarray}
where $K = q-k, D_\perp = 1/(K^2 - m_M^2), D_L = 1/(K^2 - m_D^2)$ and $D_{\lambda} = \lambda/K^4$ from the gluon propagator in the covariant gauge. We note that only the term for the gap
${\Delta_{1}}$ (henceforth referred to as ${G(q)}$) is
relevant at the Fermi surface and that it's gauge dependence multiplies
the factor 
$K^{0^2}/K^2$ which is essentially the amount by which the quarks are off mass shell. 
This is not sufficient to tame the infrared divergence from the $1/K^2$
of the gauge-fixing part of the propagator in 2+1d, in contrast to 3+1d
(see below). An estimate of the relative contributions of the electric, 
magnetic and gauge-fixing terms can now be made.
\vskip 0.3cm
\subsubsection{Gauge fixing contribution $(D_{\lambda})$}
It is the momenta on the Fermi surface that make the most contribution to the integral, therefore $|\vec{\bf{q}} - \vec{\bf{k}}| ^{2}\cong 2\mu^{2}(1-\mbox{cos}\theta)$. Setting $|\vec{\bf{q}}| - \mu \cong q_{||}$, the integration measure becomes $\  dq_{3} \mu dq_{||} d \theta$, and the continuation to Euclidean space gives
\begin{equation}
G(k_{3}) = \frac{\lambda\kappa}{2}\int \frac{dq_{3}\mu dq_{||}G(q_{3})(q_{3} - k_{3})^{2}}{q_{3}^{2} + q_{||}^{2} + G(q_{3})^{2}}\int \frac{d\theta  (1 + \mbox{cos} \theta)}{((q_{3} - k_{3})^{2} + 2\mu^{2}(1 - \mbox{cos} \theta))^2} \quad ,
\end{equation}
where $\kappa = ((N_{c}+1)/2N_{c})(g^{2}/(2\pi)^{3})$. Performing the angular integration exactly, we find
\begin{equation}
G(k_{3}) = \lambda\kappa\pi\int \frac{dq_{3}\mu dq_{||}G(q_{3})}{q_{3}^{2} + q_{||}^{2} + G(q_{3})^{2}}\left[\frac{1}{|q_{3}-k_{3}|\sqrt{(q_{3}-k_{3})^{2}+4\mu^{2}}}\right] \quad.
\end{equation}
Integrating over $q_{||}$, the pole from the quasi-particle propagator gives 
\begin{equation}
G(k_{3}) =  \lambda\kappa\mu\pi^{2}\int dq_{3}\frac{G(q_{3})}{\sqrt{q_{3}^{2}+G(q_{3})^{2}}}\left[\frac{1}{|q_{3}-k_{3}|\sqrt{(q_{3}-k_{3})^{2}+4\mu^{2}}}\right]  \quad.
\end{equation}
The region of interest is $q_{3},k_{3} \ll \mu$, since the quarks are almost on mass shell. In that case, we have
\begin{equation}
G(k_{3}) \cong  \frac{\lambda\kappa\pi^{2}}{2}\int dq_{3}\frac{G(q_{3})}{\sqrt{q_{3}^{2}+G(q_{3})^{2}}}\frac{1}{|q_{3}-k_{3}|}  \label{gauge} \quad.
\end{equation}

\subsubsection{The electric contribution ($D_{L}$)}
The angular integration yields
\begin{equation}
G(k_{3}) = \frac{\kappa\pi}{2\mu}\int \frac{dq_{3}dq_{||}G(q_{3})}{q_{3}^{2}+q_{||}^{2}+G(q_{3})^{2}}\left(\frac{\sqrt{((q_{3}-k_{3})^{2}+m_{D}^{2})^{2} + 4\mu^{2}((q_{3}-k_{3})^{2}+m_{D}^{2})}}{((q_{3}-k_{3})^{2}+m_{D}^{2})} - 1\right)  \quad.
\end{equation}
As before, the integration over $q_{||}$ picks up the pole in the diquark propagator, and the form factor is the physical regulator for the energy integral. Note that the integrand shuts off for $q_{3} \gg \mu$, but this is an unphysical regulation since we are far from the Fermi surface physics. For the physical region of $q_{3} \ll \mu$, we obtain
\begin{equation}
G(k_{3}) \cong \kappa\pi^{2}\int \frac{dq_{3} G(q_{3})}{\sqrt{q_{3}^{2}+G(q_{3})^{2}}}\frac{1}{\sqrt{(q_{3}-k_{3})^{2}+m_{D}^{2}}} \label{elec} \quad.
\end{equation}
The value of the Debye mass suggests that it is much larger than the gap, so that we may approximate $\sqrt{(q_{3}-k_{3})^{2}+m_{D}^{2}} \cong m_{D}$, with $q_{3}$ and $k_{3}$ on mass-shell being of the order of the gap. (We have checked the consistency of this approximation with the explicit solution for the gap.) The electric term is expected to be of order $\kappa/m_{D}$.

\subsubsection{The magnetic contribution ($D_{\perp}$)}
The magnetic scale is generated by Landau damping through
\begin{equation}
m_{M}^{2} = \frac{3}{4}m_{D}^{2}\frac{q_{0}-k_{0}}{|\vec{\bf{q}}-\vec{\bf{k}}|}  \quad,
\end{equation}
and the gap equation reads
\begin{equation}
G(k_{3}) = \frac{\kappa}{2}\int \frac{dq_{3}\mu dq_{||}G(q_{3})}{q_{3}^{2} + q_{||}^{2} + G(q_{3})^{2}}\int \frac{d\theta  (1 + \mbox{cos} \theta)}{(q_{3} - k_{3})^{2} + |\vec{\bf{q}}-\vec{\bf{k}}|^{2} + \frac{3}{4}m_{D}^{2}\frac{q_{3}-k_{3}}{|\vec{\bf{q}}-\vec{\bf{k}}|}} \quad.
\end{equation}
As $\vec{\bf{q}}, \vec{\bf{k}}$ are on the Fermi surface, we may write
\begin{equation}
G(k_{3}) = \frac{\kappa}{2}\int \frac{dq_{3} dq_{||}G(q_{3})}{q_{3}^{2} + q_{||}^{2} + G(q_{3})^{2}}\int \frac{\mu d\theta  (1 + \mbox{cos}\theta)}{(q_{3} - k_{3})^{2} + 2\mu^{2}(1 - \mbox{cos} \theta) + \frac{3}{4}m_{D}^{2}\frac{q_{3}-k_{3}}{\sqrt{2\mu^{2}(1 - \mbox{cos} \theta)}}} \quad ,
\end{equation}
The physics of Landau damping guides us in evaluating the angular piece. For forward scattering, $\theta$ is small, in which case the angular piece reads as follows
\begin{equation}
\int \frac{2\mu d \theta}{(q_{3} - k_{3})^{2} + \mu^{2} \theta^{2} + 3m_{D}^{2}(q_{3}- k_{3})/4\mu\theta}  \quad,
\end{equation}
where we have set $ 1 - \mbox{cos} \theta \cong \theta^{2}/2$ in the denominator and $1 + \mbox{cos} \theta \cong 2$ in the numerator. The denominator contains 3 pieces, which are, respectively, the gluon energy squared, the gluon momentum squared and the damping term. Since the scattering of BCS pairs costs little energy ( typically of the order of the gap ), the first piece may be dropped. This is equivalent to saying that the quarks are nearly on mass-shell. The validity of this approximation rests on the smallness of the gap which will be checked explicitly. 
\vskip 0.3cm
The remaining 2 terms give us a bound on the angular region in which magnetic binding of the BCS pairs is disturbed by Landau damping. Setting the terms comparable, we find
\begin{equation}
\theta_{min} = \frac{(\frac{3}{4}m_{D}^{2}|q_{3}-k_{3}|)^{1/3}}{\mu}  \quad.
\end{equation}
If $\theta < \theta_{min}$, the Landau damping is active, and is ineffective once $\theta > \theta_{min}$. Energy transfer of the order of the gap implies that $\theta_{min}$ is small if 
\begin{equation}
m_{D}^{2/3}G^{1/3} < \mu  \label{cond1} \quad.
\end{equation}
This also implies that ignoring $(q_{3} - k_{3})^{2}$ in the denominator is valid when
\begin{equation}
\frac{G}{m_{D}} < \frac{m_{D}}{\mu}   \label{cond2} \quad.
\end{equation}
It is magnetic gluon interaction between the quasiparticles that builds up the gap, as in 3+1D~\cite{Son}, so we work in the regime $\theta > \theta_{min}$. If we keep only the momentum exchange term in the denominator, we obtain for the angular integration
\begin{equation}
\frac{2}{\mu}\int_{\theta_{min}}^{\theta_{max}}\frac{d \theta}{\theta^{2}}  =  \frac{4}{{6}^{1/3}m_{D}^{2/3}|q_{3} - k_{3}|^{1/3}} \quad.
\end{equation}
 where $\mu\theta_{max} \cong (|\vec{\bf{q}}-\vec{\bf{k}}|)_{max}$ and we have picked the contribution from the lower limit. The contribution from the upper limit is subleading. The $dq_{||}$ integration yields
\begin{equation}
G(k_{3}) \cong \frac{2\pi\kappa}{6^{1/3}m_{D}^{2/3}}\int \frac{dq_{3}G(q_{3})}{\sqrt{q_{3}^{2} + G(q_{3})^{2}}}\frac{1}{|q_{3} - k_{3}|^{1/3}} \hspace{0.5in} \label{magnet} \quad.
\end{equation}
\vskip 0.3cm
Collecting Eqs.~(\ref{gauge},\ref{elec},\ref{magnet}), 
the contributions of the electric, magnetic and gauge-fixing terms are respectively
\begin{equation}
G(k_{3}) \cong \kappa\pi\left(\int \frac{dq_{3} G(q_{3})}{\sqrt{q_{3}^{2}+G(q_{3})^{2}}} \Bigg\{{\frac{\pi}{\sqrt{(q_{3}-k_{3})^{2}+m_{D}^{2}}}} + \frac{2}{6^{1/3}m_{D}^{2/3}|q_{3} - k_{3}|^{1/3}} +  \frac{\lambda\pi}{2|q_{3}-k_{3}|}\Bigg\}\right) \quad.
\end{equation}
The gauge-fixing contribution in covariant gauges is large suggesting
that the ladder plus screening expansion of the SD equation in
$\alpha/\mu$ may receive additional contributions due to the
the enhanced infrared sensitivity in 2+1d. A reorganization of the
expansion maybe needed, with perhaps an emphasis on a physical
observable from the outset. Below, we will carry estimates
of the gap in Landau gauge $\lambda=0$. The electric contribution gives
an exponentially small gap 
if $G < m_{D}$. The electric screening implies $G\sim \Lambda
e^{-m_{D}/\kappa\pi^{2}}$, 
which is exponentially suppressed due to the large value of $m_{D}/g^{2}$.
 ($\Lambda$ is the UV cutoff).

The magnetic term drives the formation of the gap since $m_{D}^{2/3}G^{1/3} < m_{D}$. The gap equation becomes
\begin{equation}
G(k_{3}) = \kappa_{*}\int \frac{dq_{3} G(q_{3})}{\sqrt{q_{3}^{2}+G(q_{3})^{2}}}  \frac{1}{|q_{3} - k_{3}|^{1/3}}  \label{magterm} \quad,
\end{equation}
where $\kappa_{*} = 2\kappa\pi/(6^{1/3}m_{D}^{2/3})$. The integral may be split into two regions, $0 < q_{3} < k_{3}$ and $ k_{3} < q_{3} < \Lambda$, where in the first, $\frac{1}{|q_{3} - k_{3}|^{1/3}} \approx \frac{1}{k_{3}^{1/3}}$, and in the second $\frac{1}{|q_{3} - k_{3}|^{1/3}} \approx \frac{1}{q_{3}^{1/3}}$.
Then, Eq.~(\ref{magterm}) may be recast as
\begin{equation}
G(k_{3}) = \kappa_{*}\left[\frac{1}{k_{3}^{1/3}}\int_{G_{0}}^{k_{3}}dq_{3} \frac{G(q_{3})}{q_{3}} + \int_{k_{3}}^{\Lambda}dq_{3}\frac{G(q_{3})}{q_{3}^{4/3}}\right]   \label{magint}\quad,
\end{equation}
with the scale $G(k_{3} = G_{0}) = G_{0}$.
Differentiating Eq.~(\ref{magint}) twice with respect to $k_{3}$, we obtain
\begin{equation}
3k_{3}^{2}G^{\prime\prime}(k_{3}) + 4k_{3} G^{\prime}(k_{3}) + \kappa_{*}\frac{G(k_{3})}{k_{3}^{1/3}} = 0  \label{BCSdiff}\quad,
\end{equation}
which has the solution
\begin{equation}
G(k_{3}) = \frac{1}{k_{3}^{1/6}}\left[c_{1}J_{-1}\left(-6\sqrt{\frac{\kappa_{*}}{3}}k_{3}^{-1/6}\right) + c_{2}Y_{-1}\left(-6\sqrt{\frac{\kappa_{*}}{3}}k_{3}^{-1/6}\right)\right]  \quad.
\end{equation}
Since the Neumann function is complex valued for positive $k_{3}$, $c_{2} = 0$ (we are solving for a real gap). Using $G(k_{3} = G_{0}) = G_{0}$ and the boundary condition $G^{\prime}(k_{3} = G_{0}) = 0$, we obtain
\begin{equation}
G(k_{3}) =
\frac{G_{0}^{7/6}}{J_{1}(6\sqrt\frac{\kappa_{*}}{3}G_{0}^{-1/6})}\frac{J_{1}(6\sqrt\frac{\kappa_*}{3}k_{3}^{-1/6})}{
k_{3}^{1/6}}  \label{magBCS} \quad,
\end{equation}
with $G_{0} = \left(\frac{\kappa_{*}}{3}\right)^{3}\left(\frac{6}{x_{0}}\right)^{6}$, where ${x_{0} = 2.405}$ is the
minimum \it{finite} \rm solution to ${J_{1}(x) = -x J_{1}'
(x)}$. ${G_{0}}$ is therefore the maximum BCS gap. Note that the condition $G_{0} < m_{D}$ is satisfied since
\begin{equation}
\frac{G_{0}}{m_{D}} \sim \left(\frac{\alpha}{\mu}\right)^{3/2}  \quad.
\end{equation}
The conditions imposed by Eqs.~(\ref{cond1}) and ~(\ref{cond2})  are also met because
\begin{equation}
\left(\frac{G_{0}}{m_{D}}\right)\left(\frac{m_{D}}{\mu}\right)^{-1} \sim \frac{\alpha}{\mu}  \quad,\quad \frac{m_{D}^{2/3}G_{0}^{1/3}}{\mu} \sim \frac{\alpha}{\mu} \quad.
\end{equation}
Therefore, the approximations made in evaluating the magnetic and electric pieces are justified. It is easy to check that the electric piece is down compared to the magnetic piece by
\begin{equation}
\frac{m_{D}^{2/3}G_{0}^{1/3}}{m_{D}} \sim \left(\frac{\alpha}{\mu}\right)^{1/2} \quad.
\end{equation}
As we work in the Landau gauge, we can repeat the preceding analysis using a
simplified propagator
\begin{equation}
D_{\mu\nu}(q) = \frac{-g_{\mu\nu}}{2}(D_{\perp}(q) + D_{L}(q)) \label{simprop}\quad,
\end{equation}
and arrive at the same equation for the gap, which in Euclidean space
is
\begin{equation}G(k) =  \frac{N_{c}+1}{2N_{c}}g^{2}\int\frac{d^{3}q}{(2\pi)^3}D(k-q)\frac{G(q)}{q_{3}^{2}+ (\mid\vec{\bf{q}}\mid - \mu)^2 + G(q)^{2}}  \quad.
\end{equation}
As expected, the integral over ${q}$ is dominated by the momenta ${
\mid\vec{\bf{q}}\mid \simeq \mu}$ and ${q_{3} \ll \mu}$.


\subsection{\underline{ Overhauser Gap equations}}
While it is possible in 3+1d
to follow the same formalism for the Overhauser pairing as in the BCS case, the method is involved in
2+1d due to the absence of ${\gamma_{5}}$, which is crucial in 
inverting ${S^{-1}}$. Chiral projectors are required to pair particles
and holes but since chirality cannot be consistently defined in 2+1d, the
arguments are more subtle. However, there is
a straightforward way and that is to use the effective action formalism developed
in~\cite{rho}.
\vskip 0.3cm The induced action in Euclidean space is
\begin{equation}
S_{\psi} = \frac{g^2}{2}\int d^{3}x\hspace{2.0mm} d^{3}y\hspace{2.0mm}
J_{\mu}^{a}(x) D_{\mu\nu}(x-y) J_{\nu}^{a}(y) + \int
d^{3}x\hspace{2.0mm}
\bar{\psi}\hspace{1.0mm}\tilde{\partial}_{\mu}\gamma_{\mu}\psi  \quad,
\end{equation}
where ${\tilde{\partial}_{\mu} = (\partial_{1}, \partial_{2},
\partial_{3} + \mu)}$ and ${J_{\mu}^{a}}$ is the usual colored current
${g\bar{\psi} \gamma_{\mu}\frac{\lambda_{a}}{2}\psi}$. We may now
Fierz the ${JJ}$ term into the relevant particle-hole channel. The
simplified propagator (Eq.~(\ref{simprop})) is used as we are working in the Landau gauge. The Fierzing factors are
\begin{equation}
\begin{array}{cr}
 Color :  -\frac{1}{2}(1-\frac{1}{N_{c}}) \,,\\ Flavor : 
\hskip 0.05cm\frac{1}{2} \,,\\ \hskip 1.5cm Spin :\,\, \frac{3}{2} \,.\\
\end{array}
\end{equation}
Following
~\cite{rho}, we introduce a hermitian bilocal field
${\Sigma(x,y)}$ to linearize the Fierzed form of the ${JJ}$ term by
using a Hubbard-Stratanovich tranformation, as
\begin{eqnarray}
\mbox{exp} \biggl(\frac{3 g^2}{16}(1-\frac{1}{N_{c}}) \int
d^{3}x\hspace{2.0mm} d^{3}y\hspace{2.0mm} [\bar{\psi}(x)\psi(y)]D(x-y)
[\bar{\psi}(y)\psi(x)]\biggr) \nonumber \\ 
= \int
d\Sigma(x,y)\hspace{2.0mm}\mbox{exp} \biggl(-S_{\Sigma} - \int
d^{3}x\hspace{2.0mm} d^{3}y\hspace{2.0mm} \bar{\psi}(x)
\Sigma(x,y)\psi(y)\biggr)  \quad,
\end{eqnarray}
with
\begin{equation}
S_{\Sigma} = \frac{4}{3g^2}(1-\frac{1}{N_{c}})^{-1} \int
d^{3}x\hspace{2.0mm}
d^{3}y\hspace{2.0mm}\frac{(\mid\Sigma(x,y)\mid)^2}{D(x-y)} \quad.
\end{equation}
The action in the quark fields is now linear and the functional
integration may be performed. Before doing so, however, we make a
simplifying ansatz for the bilocal field
\begin{equation}
\Sigma(x,y) = 2\hspace{2.0mm} \mbox{cos}\biggl[P_{\mu}\biggl(\frac{x_{\mu} +
y_{\mu}}{2}\biggr)\biggr] \sigma(x-y) = 2\hspace{2.0mm}
\mbox{cos}\biggl[P_{\mu}\biggl(\frac{x_{\mu} +
y_{\mu}}{2}\biggr)\biggr]\int\frac{d^3q}{(2\pi)^{3}}
e^{-iq\cdot(x-y)}F(q)   \quad,
\end{equation}
where ${P_{\mu} = (\vec{\bf{\mbox{P}}}_{F},0)}$ and
${\mid\vec{\bf{\mbox{P}}}_{F}\mid =
2\mu}$. ${\vec{\bf{\mbox{P}}}_{F}}$ points in the original direction
of one of the quarks. ${\Sigma}$ characterizes a standing wave of
total momentum ${2\mu}$. (The pairing is between a particle and a hole
at the opposite edges of the Fermi surface.) Introducing fermion fields
${\psi(\pm \frac{P}{2} + q)}$ as independent integration variables as
in ~\cite{dgr}, and performing the functional integration over
fermions, we obtain
\begin{equation}
\frac{3g^2}{8}(1-\frac{1}{N_{c}})\frac{S_{\Sigma}}{V_{3}} = \int
d^3x\hspace{2.0mm}\frac{\mid\sigma (x)\mid ^{2}}{D(x)} - 2g^2
\int\frac{d^3q}{(2\pi)^{3}} \mbox{ln det} \left\| \begin{array}{cr}
-\sigma\cdot Q_{+} & F \\ F & -\sigma\cdot Q_{-} \nonumber
\end{array} \right\|   \quad,
\end{equation}
where ${Q_{\pm} = [(\pm\frac{\bf{\vec{P}}}{2}+\vec{\bf{q}}),(\pm
\frac{P_{3}}{2} + q_{3} - i\mu)]}$. The determinant is over an
${(4.N_{c}.N_{f})\times (4.N_{c}.N_{f})}$  matrix. The gap equation
follows by variation and the result is
\begin{equation}
F(k) = \frac{3}{2}(N_{c}-1)g^{2} \int\frac{d^3q}{(2\pi)^{3}} D(k-q)
\frac{F(q)}{\{q_{3}+\frac{ \mid\vec{\bf{q}}\mid ^{2}}{2i\mu}\}^{2} +
q_{||}^2 + F(q)^{2}}   \quad.
\end{equation}
The measure is $d^{3}q = dq_{||}dq_{\perp}dq_{3}$, where $q_{||} = \vec{\bf{q}}\cdot {\bf{\hat{P}}}$ and ${\bf{q}_{\perp}} = {\bf{\vec{q}}} - q_{||}{\bf{\hat{P}}}$. The form factor $F(q)$ decreases rapidly as we go further from the Fermi surface, therefore, dominant contributions to the integral over $q_{||}$ come from the region where $q_{||}\sim F(q_{||})$. The gap is approximately constant in the ${\bf{q}}_{\perp}$ direction. The $q_{\perp}$ integration extends to $\Lambda_{\perp} = \sqrt{2\mu F_{0}}$, where $F_{0}$ is the maximum Overhauser gap. Thus, it picks up contributions from a larger region than $q_{||}$ does. As the particles will be placed on mass-shell, $q_{3}$ and $q_{||}$ are of the same order. The contour integration over $q_{3}$ is performed with the constraint $|{\bf{q}}_{\perp}|^{2}\leq 2\mu \epsilon_{q} \equiv 2\mu\sqrt{q_{||}^{2}+F(q_{||})^{2}}$. With these observations, the gap equation becomes
\begin{equation}
F(k_{||},k_{\perp}) = \frac{3(N_{c}-1)g^{2}}{2(2\pi)^{3}} \int dq_{||}\frac{F(q_{||})}{{\sqrt{q_{||}^{2}+F(q_{||})^{2}}}}\int_{0}^{\Lambda_{\perp}} {dq_{\perp}}D( k_{\perp}-q_{\perp}, k_{||}-q_{||} ) \quad.
\end{equation}
The electric and magnetic terms can be dealt with separately, as in the BCS case. The gap is approximately constant in the ${\bf{k}}_{\perp}$ direction, so we study the dependence on $k_{||}$ alone. \newline
With electric screening in matter, the gap equation reads
\begin{equation}
F(k_{||}) = \kappa\pi \int dq_{||}\frac{F(q_{||})}{{\sqrt{q_{||}^{2}+F(q_{||})^{2}}}}\frac{1}{\sqrt{m_{D}^{2} + (k_{||}-q_{||})^{2}}}\mbox{tan}^{-1}\left(\frac{\Lambda_{\perp}}{\sqrt{m_{D}^{2} + (k_{||}-q_{||})^{2}}}\right)  \,,
\end{equation}
where $\kappa = 3(N_{c}-1)g^{2}/4(2\pi)^{3}$. 
\vskip 0.2cm
In case of Landau damping of the magnetic mode, the gap equation is
\begin{equation}
F(k_{||}) \cong \kappa\pi \int dq_{||}\frac{F(q_{||})}{{\sqrt{q_{||}^{2}+F(q_{||})^{2}}}}\,\frac{c}{(\frac{3}{4}m_{D}^{2}|k_{||}-q_{||}|)^{1/3}} \label{magapprox}  \quad,
\end{equation}
where $c = (6\,\mbox{ln}2 + 2\sqrt{3}\pi)/18$. In writing Eq~(\ref{magapprox}), we have expanded in $(\frac{3}{4}m_{D}^{2}|k_{||}-q_{||}|)^{1/3}/\Lambda_{\perp}$ and retained only the leading term. This ratio is $N_{c}$ dependent and the expansion turns out to be valid only at sufficiently large $N_{c}$ (see Eq~(\ref{magvalid})). Noting that $m_{D}$ is typically much larger than the gap, the magnetic binding will drive the formation of the gap. Proceeding as in the BCS case, we convert the integral equation into a differential one,
\begin{equation}
3k_{||}^{2}F^{\prime\prime}(k_{||}) + 4k_{||} F^{\prime}(k_{||}) + \kappa_{*}\frac{F(k_{||})}{k_{||}^{1/3}} = 0 \label{Ovdiff}  \quad,
\end{equation}
where $\kappa_{*} = (c\pi\kappa)/(\frac{3}{4}m_{D}^{2})^{1/3}$. Eq.~(\ref{Ovdiff}) is analogous to Eq~(\ref{BCSdiff}) obtained for the BCS gap. The solution follows 
\begin{equation}
F(k_{||}) =
\frac{F_{0}^{7/6}}{J_{1}(6\sqrt\frac{\kappa_{*}}{3}F_{0}^{-1/6})}\frac{J_{1}(6\sqrt\frac{\kappa_{*}}{3}k_{||}^{-1/6})}{
k_{||}^{1/6}}   \quad,
\end{equation}

with $F_{0} = \left(\frac{\kappa_{*}}{3}\right)^{3}\left(\frac{6}{x_{0}}\right)^{6}$, the maximum Overhauser gap. Note that the condition $F_{0} < m_{D}$ is satisfied since $F_{0}/m_{D} \sim (\alpha/\mu)^{3/2}$. As in the BCS case, $F_{0}/\mu \sim (\alpha/\mu)^{2}$. These observations justify the neglect of electric screening effects. The validity of the approximation in Eq.~(\ref{magapprox}) implies that the ratio
\begin{equation}
\frac{(\frac{3}{4}m_{D}^{2}F_{0})^{1/3}}{\Lambda_{\perp}} =
\left(\frac{2\pi x_{0}^{2}}{6c(N_{c}-1)}\right)^{1/2} \cong \left(\frac{7.24}{(N_{c}-1)}\right)^{1/2}\label{magvalid}  \,.
\end{equation}
should be small. So long as $N_{c}$ is sufficiently large, the approximation and the resulting estimate of the Overhauser gap can be trusted. If $N_c$ does not meet this condition, the binding is disturbed throughout the pairing region by Landau damping. With this caveat, we can make a comparison of the maximum BCS and Overhauser gaps ($F_{0}$ and $G_{0}$). We find
\begin{equation}
\frac{F_{0}}{G_{0}} \cong 2\left(N_{c}\frac{(1 - 1/N_{c})}{(1 + 1/N_{c})}\right)^{3} \,,
\end{equation}
which shows that in 2+1d, the Overhauser gap is larger than the BCS gap for sufficiently large $N_{c}$. Finally, an assessment of the energy budget for the two competing mechanisms yields
\begin{equation}
\frac{\epsilon_{OV}}{\epsilon_{BCS}} = \frac{(\Lambda_{\perp}F_{0})F_{0}}{(\mu G_{0})G_{0}} \cong \left(\frac{F_{0}}{G_{0}}\right)^{2}\left(\frac{F_{0}}{\mu}\right)^{1/2}  \,\,.
\end{equation}


 
\section{Conclusions}
In considering BCS and Overhauser pairing in 2+1 dimensions, we
have argued for the form of the condensates that we have chosen to study. The
gap equations are written down and treated with the effects of
electric screening and Landau damping in matter. For the case of 2+1d,
and to 
leading order in perturbation theory, we have derived modifications to 
the gluon propagator from screening by quarks, using the imaginary time formalism.
The standard Nambu-Gorkov formalism is more subtle for 
the Overhauser case due to the absence of ${\gamma_{5}}$ 
and the effective action method is applied instead. A comparison
of the BCS and Overhauser gap in Landau gauge,
shows that the latter is preferred with increasing $N_{c}$. 
The energy budgeting shows that BCS pairing is negatively affected by
the reduction of 
available phase space for pairing in the lower dimension. 
The gaps themselves are small, being of order $(\alpha^{2}/\mu)$,
although the enhanced infrared sensitivity in 2+1d prevents us from
a definitive conclusion on their gauge independence. 
Also, the coupling does not run in 2+1d, so it is not possible 
to set a perturbative scale, and consequently to say when 
$\alpha/\mu$ is small. Our perturbative analysis assumes that 
we can take $\alpha/\mu$ small. It is interesting to note 
that unlike the case of 3+1d or 1+1d, the BCS and Overhauser gaps 
show power law dependence ($g^{4}/\mu$) in Landau gauge. Therefore, they are more 
sensitive to higher order corrections from perturbation theory than 
their corresponding forms in 3+1d.  A numerical analysis of the gap 
equations in 2+1d would complement our theoretical approach, and would 
also serve to test the validity of some of the approximations. Finally, it would 
be interesting to consider the effects of the crystalline
superconducting 
phase discussed recently by others ~\cite{KRISHNA,raja} as a competing 
instability in this space-time dimension.
\vskip 0.5cm
\section*{Acknowledgements}
We are grateful to Thomas Sch\"afer for useful discussions. This work was supported by the US-DOE grant DE-FG-88ER40388.
\vskip 1cm
\section*{Note}
While writing our paper, we were informed that a related analysis for the BCS case was being performed by V. A. Miransky, G. W. Semenoff, I. A. Shovkovy and L. C. R. Wijewardhana.

\begin{flushleft}

\end{flushleft}

\end{document}